%% file: top_dil_prd-rc_v2.tex
\documentclass[aps,prl,twocolumn,superscriptaddress,showpacs]{revtex4}

\usepackage{graphicx}

\usepackage{tabularx}
\usepackage{microtype}

\newcommand{\ttbar}{ t \bar{t} }
\newcommand{\Ht}{H_T}
\newcommand{\pt}{p_T}
\newcommand{\met}{{\nobreak \not\!\!E_{\rm T}}}

\newcommand{\WG}{W\gamma}
\newcommand{\pythia}{{\sc pythia}}
\newcommand{\alpgen}{{\sc alpgen}}
\newcommand{\MCFM}{{\sc mcfm}}
\newcommand{\SecVtx}{{\sc secvtx}}
\newcommand{\DY}{Z/\gamma ^*}
\newcommand{\MSTW}{{MSTW}}

\begin{document}

\title{ 
Measurement of the top-quark pair production cross-section in  events
with  two leptons and  bottom-quark jets using the full CDF data set
}

\date{\today}

\input{November2012_Authors}
\begin{abstract}
We present a measurement of the top-quark pair production cross-section in proton-antiproton collisions 
at $\sqrt{s}$=1.96 TeV. The data were collected at the Fermilab Tevatron by the CDF~II detector 
and correspond to an integrated luminosity of 8.8 fb$^{-1}$, representing the complete CDF Run~II data set.  
We select events consistent with the production of top-quark pairs by requiring  the presence of two 
reconstructed leptons, an imbalance in the total event transverse momentum, and jets. 
At least one jet is required to be identified as consistent with the fragmentation of a bottom quark using 
a secondary-vertex-finding algorithm. 
The 246 candidate events are estimated to have a signal purity of 91\%. 
We measure a cross section of $\sigma_{\ttbar}$ = 7.09 $\pm$ 0.84~pb, assuming a top-quark mass of 172.5~GeV/$c^{2}$. 
The results are consistent with the standard model as predicted by next-to-leading-order calculations.
\end{abstract}


\maketitle

%


Beginning with the discovery of the top quark ($t$) in 1995~\cite{top_cdf, top_d0}, the CDF and D0
experiments at Fermilab have studied its production, decays, and intrinsic properties~\cite{TOPMASS, HELICITY, RATIO, D0_XSEC_COMBO, CDF_DIL_2p8, CDF_LJ_XSEC}.  
This Letter continues that rich program by reporting the first top-antitop quark pair ($t\bar{t}$) production cross-section measurement by a Tevatron experiment that utilizes the complete Run~II data set.
Studies of the top quark provide both measurements of standard model (SM)
parameters~\cite{Precision_EW} and probes of non-SM particles or
interactions~\cite{theory}.
Top-quark pairs are produced in proton-antiproton collisions at $\rm \sqrt{s}=1.96$ TeV by the Fermilab Tevatron.
We select events for this measurement if both leptons 
in the decay chain $\ttbar \rightarrow (W^+b)(W^-\bar{b})\rightarrow
(\ell^+{\nu}_lb)(\ell^-\bar{\nu}_l\bar{b})$  are identified.
Only reconstructed electron or muon candidates are selected as leptons.  Hadronic decays of tau leptons are not considered.
The signal yield is measured as the number of selected events in the data after subtraction of the background expectation from 
other SM sources, and the cross section is measured by correcting the signal
yield for acceptance, efficiency, and luminosity.  This analysis uses the full CDF Run~II data set collected between March 2002 and September 2011,
which corresponds to 8.8~fb$^{-1}$ of integrated luminosity after data-quality requirements are imposed.
The result supersedes a previous  analysis~\cite{CDF_DIL_2p8}  
by  exploiting a three-fold increase in data set and improved $\ttbar$ signal-to-background ratio.
Improved sample purity is obtained by requiring the presence of jets consistent with the fragmentation of $b$~quarks ($b$-tagged jets) from the top quark decay.


The CDF~II detector is a solenoidal spectrometer surrounded by a sampling calorimeter and muon detectors~\cite{CDFII}. 
%
%
CDF uses a cylindrical coordinate system with the origin at the center of 
the detector and the +$z$ direction defined by the proton beam. $\theta$ 
is the polar angle with respect to the $z$-axis and $\phi$ is the azimuthal angle.

The events were selected for analysis during data taking with an inclusive selection that 
required the presence of an electron (a muon) with $E_T$ $> \:$18~GeV  ($\pt$ $> \:$ 18~GeV/$c$). 
The transverse energy and momentum are defined as $E_T$ =
$E\sin(\theta)$ and $\pt$ = $p\sin(\theta)$ where $E$ is the energy
measured in the calorimeter and $p$ is the momentum measured by the
tracking system.
In the offline analysis, we select events that contain at least one isolated~\cite{isolated} electron (muon) 
with $E_T$ $> \:$ 20~GeV ($\pt$ $> \:$ 20~GeV/$c$).  
We additionally require the presence of a second lepton with the same energy requirements, 
but without isolation requirements.  Events with more than two reconstructed leptons are rejected.

The neutrinos from dilepton top-quark pair decays escape detection, 
so signal events are expected to produce a large imbalance in the event total transverse-energy ($\met$)~\cite{MET} compared 
to other SM processes containing two leptons.
We require $\met$ $>$ 25~GeV to reduce contamination from processes that do
not involve neutrinos from vector-boson decays.
Events in which $\met$ originates from instrumental effects typically feature a
small angle between the direction of a lepton or jet and the direction of $\vec{\met}$.  
If this angle is smaller than 20$\rm ^o$, we require $\met >$ 50~GeV to reject these backgrounds. 
To specifically reject events from $\DY$ production, we require high $\met$ significance~\cite{CDF_DIL_2p8} if the identified leptons 
have the same flavor and dilepton mass consistent with the $Z$ resonance.
We also require the dilepton mass to be larger than 5~GeV/$c^{2}$  to 
remove events from low-mass dimuon resonances.  The resulting sample is referred to as events meeting 
the {\it dilepton selection}.  Jets are identified in the laboratory frame using a modified cone
algorithm~\cite{jetclu}, and are defined as having $E_T$ $> \:$15~GeV and pseudorapidity
in the lab frame satisfying $|\eta| < 2.5$. Events satisfying the dilepton selection 
that contain exactly zero or one jet are used as  control samples for background estimation. 
The {\it pretag sample} contains events passing the dilepton selection with at
least two jets, 
summed transverse energy over all particles ($\Ht$) satisfying $\Ht$ $>$ 200~GeV, 
and whose two leptons are of opposite electric charge.  
The data sample corresponds to an integrated luminosity of 9.1~fb$^{-1}$, slightly higher 
than the signal sample, because the detector quality requirements 
for $b$ tagging are not imposed.  The pretag sample is used to validate the signal and background models.  
We measure the $t\bar{t}$ production cross-section using the {\it tag sample}, 
which is the subset of pretag events in which at least one of the jets in the event is $b$ tagged by the \SecVtx{} algorithm~\cite{secvtx}.

The lifetime of $B$ hadrons is approximately 1.5 picosecond, so relativistic $B$ hadrons produced 
in collisions at the Tevatron can travel on the order of 450 micrometers from the primary 
interaction-point (primary vertex) before decaying.  We use charged-particle tracks to reconstruct the 
primary vertex and secondary decay-vertices.  We then compute the two-dimensional 
displacement of the secondary vertex from the primary vertex projected along the jet direction in the plane transverse to the beam ($L_{2D}$).
A jet is considered $b$-tagged by the \SecVtx{} algorithm if $L_{2D}$ and its 
uncertainty ($\sigma$) satisfy the significance $L_{2D}/\sigma>7.5$.  
Jets with $L_{2D}/\sigma<-7.5$ are not topologically consistent with $B$-hadron decays, 
but are used to estimate the false-tag rate due to instrumental sources~\cite{secvtx}.


Selection efficiency for $t\bar{t}$ events is estimated using the \pythia{}
Monte Carlo  event-generator~\cite{pythia} combined with a detailed simulation 
of the CDF II detector~\cite{GEANT3}.
The $t\bar{t}$ signal is simulated assuming a top-quark mass of 172.5~GeV/$c^2$ and only contains events
 in which both $W$ bosons produced by the decay of the top quarks subsequently decay into a
 charged lepton ($e$, $\mu$, $\tau$) and a neutrino.
 Only simulated events with a 
 primary vertex reconstructed within $\rm \pm60$~cm of the nominal CDF detector 
 center are retained.  This requirement has an efficiency of [97.47 $\pm$
 0.02(stat)]\% of the
  full CDF luminous region.  The total acceptance for the $b$-tagged (pretag) dilepton signal events 
  is [0.461 $\pm$ 0.003(stat)]\% ([0.756 $\pm$ 0.004(stat)]\%), including the branching fraction to leptons.
%
%
This acceptance must be corrected to account
for the efficiency of the inclusive lepton triggers, which are measured using data samples
selected by an independent set of triggering criteria and are in the range 85-95\%.
The simulation is also corrected for imperfect modeling of the lepton identification
efficiencies by measuring these efficiencies using $\DY \rightarrow \ell\ell$ events in the data, where one
lepton ($\ell$ = $e$ or $\mu$) is fully identified and the other is used for the efficiency measurement. 
We use these efficiencies to derive multiplicative correction factors (in the range 0.8 - 1.0)
to apply to the simulated efficiencies.  
We correct for the difference in efficiency for the $b$-tagging algorithm between data 
and simulated samples by using a multiplicative correction factor, $S_b = 0.96 \pm 0.05$. 
This correction accounts for the differences between properties of jets 
in $\ttbar$ events and jets in the $b$-tagging calibration sample~\cite{secvtx}.


The relevant background processes yielding prompt lepton pairs are  diboson ($WW$, $WZ$, and $ZZ$) 
production and $\DY$ production. Processes in which a photon or hadronic fragmentation are 
identified as a lepton are also considered, such as $\WG$ and $W$ boson production in association with multiple jets.
The signal sample contamination is predicted to predominantly comprise $\DY$ and 
$W$+jets production processes, so their normalizations are estimated using data samples enriched in these processes. 
The contamination for the remaining backgrounds is predicted using the same detector simulation and corrections used for signal.

Diboson production is simulated with  \pythia{}, normalized to the production
cross-sections from the next-to-leading order
calculations using {\rm MCFM}~\cite{NLO} and \MSTW 2008~\cite{PDF} parton distribution functions (PDF). 
The predicted cross sections are 
$\sigma _{\it WW}$ = 11.34 $\pm$ 0.68~pb, $\sigma _{\it WZ}$ = 3.47 $\pm$ 0.21~pb,
and $\sigma _{\it ZZ}$ = 3.62 $\pm$ 0.22~pb~\cite{tevnphwg}. 
 The $\DY \rightarrow \ell^+\ell^-$ production is simulated using the
\alpgen+\pythia{} event generator~\cite{alpgen}. 
The $\DY \rightarrow e^+e^-$,  $\mu^+\mu^-$ samples, which are only selected
due to instrumental mismeasurements, are normalized in a data derived process.  
The process $\DY\rightarrow\tau^+\tau^-$  has significant $\met$ from neutrinos,
and is treated separately; it is normalized to the \alpgen{} production rate, corrected for next-to-leading-order contributions~\cite{CDF_DIL_2p8}.
 The $\WG$ decays are simulated with the {\sc baur} event generator~\cite{baur}, 
assuming a leading-order production cross-section of $\sigma _{\WG}$ = 32 $\pm$ 3~pb and 
correcting for higher-order effects~\cite{baur2}.  This process is observed to be 
relevant in low jet-multiplicity control samples, and negligible in the signal sample.

The $WW$ and $\DY \rightarrow \tau\tau$ jet multiplicity spectra 
are corrected to account for discrepancies observed between data and simulation
in $Z$ boson decays, using jet-multiplicity-dependent correction factors. 
The corrections are applied to processes in which jets are produced by initial-state radiation, 
rather than from final-state partons in the hard scattering. 
The uncertainties on the acceptances of the simulated background processes come from the 
convolution of the uncertainties due to finite simulation sample-size and uncertainties
on the jet-multiplicity correction factors, lepton identification,
 and jet energy scale~\cite{JES}.

The sample contamination from $\DY$ to $ee$ and $\mu\mu$ decays with instrumental missing energy is estimated 
in a data sample in which the dilepton mass for all events is consistent with the
$Z$ resonance, 
but all other selection criteria are that of the pretag sample. 
We subtract the contributions from other processes, and then 
extrapolate the observed rate outside of the resonance region by using
simulated samples,
 independently for each lepton type and jet-multiplicity.
The uncertainty on this background contribution is dominated by the limited number of $\DY$ data events 
with high $\met$ used to normalize the overall prediction, from the finite size
of the  $\DY$ events that meet the selection,
 and from the uncertainty on the jet-energy scale.  
The normalization of the \mbox{$\DY + u,d,s,${\textls[-500]{ }}\textsl{g}} event yield in the tagged sample is determined by 
applying the scalings determined in the pretag sample, and applying false $b$-tag rates~\cite{secvtx} as 
weights to the \alpgen+\pythia{} events.  We obtain the $\DY+b,c$ event yield normalization in the tagged sample 
by requiring events with dilepton mass consistent with the $Z$ resonance, but all other selection criteria as that of the tag sample.  
After subtracting the estimated \mbox{$\DY + u,d,s,${\textls[-500]{ }}\textsl{g}} component and other backgrounds, the multiplicative 
heavy-flavor-specific $\DY$ normalization corrections are found to be $1.8 \pm 0.1.$

We estimate a small contribution to the sample of events with one electron 
and one muon from  $\DY$ $\rightarrow$  $\mu\mu$
events,  in which bremsstrahlung associated with one muon mimics an electron signature.
These events are described using the $\DY\rightarrow \mu\mu$ simulation sample.
%
%
The background from jets misidentified as leptons is estimated by using data events 
with exactly one identified lepton and additional lepton-like  
candidates that satisfy less restrictive identification criteria (called the ``W+jet'' sample,
although other sources of misidentified leptons contribute to this sample as well).  
%
%
%
The probability that a lepton-like candidate is reconstructed as a lepton is parametrized in terms 
of the candidate's transverse energy and isolation, and measured in large QCD dijet dominated samples 
triggered by the presence of at least one jet of $E_{T}>$ 50 GeV (called the ``jet samples'')~\cite{CDF_DIL_2p8}.
Misidentified leptons are modeled by applying
these probabilities as weights to the events in the W+jet sample with only one high
transverse-energy reconstructed lepton and a second electron-like or muon-like
candidate. To remove events with two good leptons from this sample, the lepton-like candidate 
is required to fail at least one lepton identification requirement.
%
%
The uncertainty on the misidentified-lepton background model is dominated by the differences observed between
identification rates determined in jet samples triggered by jets with $E_T$ greater than 20, 50, 70, or 100~GeV.

A common systematic uncertainty for signal and simulated background estimates comes from the uncertainty on the lepton identification
correction factors, which is measured to be 2.2\%. The 3.3\% uncertainty due to the jet-energy scale affects all simulated samples, and is estimated by varying the jet-energy corrections by $\pm 1 \sigma$ of their systematic uncertainty and measuring the shift in signal and background acceptance.
 We consider several other sources of systematic uncertainties predominantly affecting the signal efficiency: 
 difference in $\ttbar$ modeling by various simulation generators, 
 simulation of initial- and final-state radiation, color reconnection, and PDF~\cite{PDF} uncertainty.
These are determined by comparing the uncorrected simulation acceptance of the 
default $\ttbar$ {\sc pythia} sample to specialized simulation samples.  
The uncertainty due to each of these sources is estimated to be
less than 2.0\%. 
The systematic uncertainty due to the $b$-tagging efficiency correction is
5.0\%, dominated by the light-flavor modeling.
All simulated backgrounds have uncertainty due to the jet-multiplicity correction factor. 
Uncorrelated sources of systematic uncertainties affecting individual backgrounds 
include the 30\% systematic uncertainty on the misidentified lepton contamination 
and individual theoretical uncertainties, ranging from 2\% to 10\%, on the production cross-sections of diboson and $\DY \rightarrow \tau\tau$ processes.
Each of these effects contributes to only a small fraction of the resulting 2.1\% (7.1\%) background systematic uncertainty
for the $b$-tagged (pretag) sample.
Table \ref{tab:syst} summarizes the systematic uncertainties that affect 
the signal acceptance and background model~\cite{CDF_DIL_2p8}.


\begin{table}[htbp]
\caption{Systematic uncertainties for the pretag and \mbox{$b$-tagged} samples. 
The total systematic uncertainty is the sum in quadrature of each independent contribution.} \label{tab:sys}

\begin{center}

\begin{tabular*}{0.46\textwidth}{ l c c}
\hline \hline
\multicolumn{1}{c}{Source} & \multicolumn{2}{c}{Uncertainty (\%)} \\
               			   		 &  \phantom{X}Pretag\phantom{X} & \phantom{X}$b$-tagged\phantom{X}\\		
\hline 
Lepton identification efficiency	             &  2.2		&  2.2   \\
Jet energy scale           	    	             &  3.3 	&  3.3   \\
Simulated event generator                      &  1.9 	&  1.9   \\
Initial- and final-state radiation\phantom{X}  &  1.3 	&  1.3   \\
Color reconnection         	                   &  1.2  	&  1.2   \\
PDF                      		                   &  0.6 	&  0.6   \\
$b$-tagging                	                   &   - 		&  5.0   \\
Background model           	                   &  7.1		&  2.1   \\
\hline
Total systematic uncertainty                   &  8.6   &  7.2   \\
\hline\hline
\end{tabular*}
\end{center}
\label{tab:syst}
\end{table}

\begin{figure}[htbp]
 \begin{center}
    \includegraphics[width=0.45\textwidth,  clip]{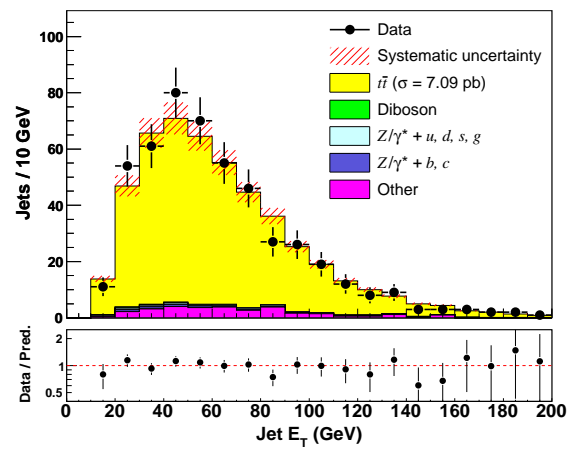}
 \end{center}
  \caption{
           Distribution of jet $E_{T}$ values for the two jets with largest $E_{T}$ in each event (black points) for the signal sample.  
	   The histogram represents the sum of the signal and background estimates, 
	   where the signal is normalized to the observed cross section. 
     The hatched area is the total uncertainty on the sum of the signal and background predictions.
     The lower panel shows the observed data yield divided by the predicted yield. }
  \label{fig:JetEt}
\end{figure}

The expected and observed background events that are $b$-tagged in the 1-jet sample are used as a control sample. 
The final sample of events with two or more jets passing all candidate selection 
criteria is given in Table \ref{tab:btag}. 
 The signal purity in the tag sample is 91\%, which can be compared to the 73\% achieved in the pretag sample.  
 In Fig.~\ref{fig:JetEt}, we present the jet $E_T$ spectrum for the leading two jets in events with at least two jets, and at least one $b$-tag.  
 The signal yield in the figure is normalized to the measured cross section, and the shape of the distribution is well described by the prediction.

\begin{table}[htbp]
\caption{ Estimated number of background and $\ttbar$ signal events in the $b$-tagged sample, 
which corresponds to an integrated luminosity of 8.8~fb$^{-1}$.  The observed event yields are 
compared with the total SM expectation for both the 1-jet and signal samples.   
The quoted uncertainties are the quadratic sum of the statistical and systematic uncertainties in each row.  In the right column, \lq\lq$\Ht$+OS\rq\rq 
refers to the requirements that events contain leptons with opposite electric charge and satisfy
$\Ht > 200$ ~GeV. These requirements are not applied to the events in the left column.}

\begin{center}
\renewcommand{\tabcolsep}{0.30mm}
\begin{tabular*}{0.49\textwidth}{@{\extracolsep{\fill}} l rcl rcl }
\hline 
\hline
 		                                   	 & \multicolumn{3}{c}{1 jet}               & \multicolumn{3}{c}{$\ge 2$ jets ($\Ht$+OS)} \\
Source                                   & \multicolumn{3}{c}{(Validation region)} & \multicolumn{3}{c}{(Signal region)}   \\
                                         \hline
$WW$                                     &~~~~   0.8     &   $\pm$   &     0.2  ~~~&~~~   0.6     &    $\pm$ &    0.2 ~~~ \\
$WZ$                                     &~~~~   0.2     &   $\pm$   &     0.0  ~~~&~~~   0.1     &    $\pm$ &    0.0 ~~~ \\
$ZZ$                                     &~~~~   0.1     &   $\pm$   &     0.0  ~~~&~~~   0.3     &    $\pm$ &    0.1 ~~~ \\
$\DY+u,d,s,${\textls[-500]{ }}\textsl{g} &~~~~   2.1     &   $\pm$   &     0.2  ~~~&~~~   2.8     &    $\pm$ &    0.3 ~~~ \\
$\DY+b,c$                                &~~~~   1.8     &   $\pm$   &     0.2  ~~~&~~~   2.5     &    $\pm$ &    0.2 ~~~ \\
Other		                                 &~~~~   1.9     &   $\pm$   &     0.7  ~~~&~~~    16     &    $\pm$ &      5 ~~~ \\
\hline                                                                                                                   
Total background                         &~~~~   6.9     &   $\pm$   &     0.9  ~~~&~~~    22     &    $\pm$ &      5 ~~~ \\
$t\bar{t}$ ($\sigma = 7.09$ pb)          &~~~~  20.2     &   $\pm$   &     1.4  ~~~&~~~   224     &    $\pm$ &     15 ~~~ \\
\hline                                                                                                                     
Total SM expectation                     &~~~~  27.1     &   $\pm$   &     2.2  ~~~&~~~   246     &    $\pm$ &     20 ~~~ \\
Observed                                 &           \multicolumn{3}{c}{~~29}      ~~   &     \multicolumn{3}{c}{246} ~          \\ 
\hline\hline
\end{tabular*}
\end{center}
\label{tab:btag}
\end{table}

The measured cross section is calculated as
\begin{equation}\label{xsec}
{ \sigma _{\ttbar} = \rm \frac{N_{obs}-N_{bkg}}{\sum _i
\mathcal{A}_i  \mathcal{L}_i}},
 \end{equation}
where N$\rm _{obs}$ is the number of dilepton candidate events,
N$\rm _{bkg}$ is the total number of expected background events, and the denominator is the
weighted sum of the corrected acceptance for each class of events grouped by lepton reconstruction~\cite{CDF_DIL_2p8}.
We multiply $\mathcal{A}_i$ by the integrated luminosity corresponding to the reconstruction class $\mathcal{L}_i$. 
Various values of integrated luminosities are used because the identification of events as belonging to
each lepton class requires different CDF subdetectors to be fully
functional. The total denominator for the $b$-tagged  events is 
31.60 $\pm$ 0.19~pb$^{-1}$. 


For the $b$-tagged $\ttbar$ dilepton sample, we measure a cross section of
$\sigma_{\ttbar}$ = 7.09 $\pm$ 0.49(stat) $\pm$ 0.52(syst) $\pm$ 0.43(lumi)~pb =
7.09 $\pm$ 0.84~pb with the 246 signal candidate events.  
The systematic uncertainty is the convolution of the acceptance and the background uncertainties shown in Table~\ref{tab:syst}.  The 6\% luminosity uncertainty is kept separate~\cite{lum}.
The results presented here are consistent with the best recent predictions from next-to-leading order theoretical calculations~\cite{theory}, 
and with previous D0 and CDF publications~\cite{D0_XSEC_COMBO, CDF_DIL_2p8}.
 The current data sample corresponds to an integrated luminosity three times greater than
 that of the previous publication~\cite{CDF_DIL_2p8},
   producing a result with a statistical uncertainty which is smaller than the systematic uncertainty.
The use of $b$-jet identification further improves the signal purity from 73\%
to 91\%, and the total uncertainty of the measurement has been improved from 1.04 pb in the previous CDF publication to the current value of 0.84 pb. 

In conclusion, we have measured the production cross-section of top-quark pairs at the Tevatron, 
using the full CDF Run~II data set.  This measurement offers a robust addition to global combined measurements of the top-quark production cross-section, which can then be used as constraints to theoretical calculations and limits on non-SM contributions in the top-quark sector.

We thank the Fermilab staff and the technical staffs of the participating institutions for their vital contributions. This work was supported by the U.S. Department of Energy and National Science Foundation; the Italian Istituto Nazionale di Fisica Nucleare; the Ministry of Education, Culture, Sports, Science and Technology of Japan; the Natural Sciences and Engineering Research Council of Canada; the National Science Council of the Republic of China; the Swiss National Science Foundation; the A.P. Sloan Foundation; the Bundesministerium f\"ur Bildung und Forschung, Germany; the Korean World Class University Program, the National Research Foundation of Korea; the Science and Technology Facilities Council and the Royal Society, UK; the Russian Foundation for Basic Research; the Ministerio de Ciencia e Innovaci\'{o}n, and Programa Consolider-Ingenio 2010, Spain; the Slovak R\&D Agency; the Academy of Finland; the Australian Research Council (ARC); and the EU community Marie Curie Fellowship contract 302103.


\end{document}

%% file: November2012_Authors.tex
\affiliation{Institute of Physics, Academia Sinica, Taipei, Taiwan 11529, Republic of China}
\affiliation{Argonne National Laboratory, Argonne, Illinois 60439, USA}
\affiliation{University of Athens, 157 71 Athens, Greece}
\affiliation{Institut de Fisica d'Altes Energies, ICREA, Universitat Autonoma de Barcelona, E-08193, Bellaterra (Barcelona), Spain}
\affiliation{Baylor University, Waco, Texas 76798, USA}
\affiliation{Istituto Nazionale di Fisica Nucleare Bologna, $^{ee}$University of Bologna, I-40127 Bologna, Italy}
\affiliation{University of California, Davis, Davis, California 95616, USA}
\affiliation{University of California, Los Angeles, Los Angeles, California 90024, USA}
\affiliation{Instituto de Fisica de Cantabria, CSIC-University of Cantabria, 39005 Santander, Spain}
\affiliation{Carnegie Mellon University, Pittsburgh, Pennsylvania 15213, USA}
\affiliation{Enrico Fermi Institute, University of Chicago, Chicago, Illinois 60637, USA}
\affiliation{Comenius University, 842 48 Bratislava, Slovakia; Institute of Experimental Physics, 040 01 Kosice, Slovakia}
\affiliation{Joint Institute for Nuclear Research, RU-141980 Dubna, Russia}
\affiliation{Duke University, Durham, North Carolina 27708, USA}
\affiliation{Fermi National Accelerator Laboratory, Batavia, Illinois 60510, USA}
\affiliation{University of Florida, Gainesville, Florida 32611, USA}
\affiliation{Laboratori Nazionali di Frascati, Istituto Nazionale di Fisica Nucleare, I-00044 Frascati, Italy}
\affiliation{University of Geneva, CH-1211 Geneva 4, Switzerland}
\affiliation{Glasgow University, Glasgow G12 8QQ, United Kingdom}
\affiliation{Harvard University, Cambridge, Massachusetts 02138, USA}
\affiliation{Division of High Energy Physics, Department of Physics, University of Helsinki and Helsinki Institute of Physics, FIN-00014, Helsinki, Finland}
\affiliation{University of Illinois, Urbana, Illinois 61801, USA}
\affiliation{The Johns Hopkins University, Baltimore, Maryland 21218, USA}
\affiliation{Institut f\"{u}r Experimentelle Kernphysik, Karlsruhe Institute of Technology, D-76131 Karlsruhe, Germany}
\affiliation{Center for High Energy Physics: Kyungpook National University, Daegu 702-701, Korea; Seoul National University, Seoul 151-742, Korea; Sungkyunkwan University, Suwon 440-746, Korea; Korea Institute of Science and Technology Information, Daejeon 305-806, Korea; Chonnam National University, Gwangju 500-757, Korea; Chonbuk National University, Jeonju 561-756, Korea; Ewha Womans University, Seoul, 120-750, Korea}
\affiliation{Ernest Orlando Lawrence Berkeley National Laboratory, Berkeley, California 94720, USA}
\affiliation{University of Liverpool, Liverpool L69 7ZE, United Kingdom}
\affiliation{University College London, London WC1E 6BT, United Kingdom}
\affiliation{Centro de Investigaciones Energeticas Medioambientales y Tecnologicas, E-28040 Madrid, Spain}
\affiliation{Massachusetts Institute of Technology, Cambridge, Massachusetts 02139, USA}
\affiliation{Institute of Particle Physics: McGill University, Montr\'{e}al, Qu\'{e}bec H3A~2T8, Canada; Simon Fraser University, Burnaby, British Columbia V5A~1S6, Canada; University of Toronto, Toronto, Ontario M5S~1A7, Canada; and TRIUMF, Vancouver, British Columbia V6T~2A3, Canada}
\affiliation{University of Michigan, Ann Arbor, Michigan 48109, USA}
\affiliation{Michigan State University, East Lansing, Michigan 48824, USA}
\affiliation{Institution for Theoretical and Experimental Physics, ITEP, Moscow 117259, Russia}
\affiliation{University of New Mexico, Albuquerque, New Mexico 87131, USA}
\affiliation{The Ohio State University, Columbus, Ohio 43210, USA}
\affiliation{Okayama University, Okayama 700-8530, Japan}
\affiliation{Osaka City University, Osaka 588, Japan}
\affiliation{University of Oxford, Oxford OX1 3RH, United Kingdom}
\affiliation{Istituto Nazionale di Fisica Nucleare, Sezione di Padova-Trento, $^{ff}$University of Padova, I-35131 Padova, Italy}
\affiliation{University of Pennsylvania, Philadelphia, Pennsylvania 19104, USA}
\affiliation{Istituto Nazionale di Fisica Nucleare Pisa, $^{gg}$University of Pisa, $^{hh}$University of Siena and $^{ii}$Scuola Normale Superiore, I-56127 Pisa, Italy, $^{mm}$INFN Pavia and University of Pavia, I-27100 Pavia, Italy}
\affiliation{University of Pittsburgh, Pittsburgh, Pennsylvania 15260, USA}
\affiliation{Purdue University, West Lafayette, Indiana 47907, USA}
\affiliation{University of Rochester, Rochester, New York 14627, USA}
\affiliation{The Rockefeller University, New York, New York 10065, USA}
\affiliation{Istituto Nazionale di Fisica Nucleare, Sezione di Roma 1, $^{jj}$Sapienza Universit\`{a} di Roma, I-00185 Roma, Italy}
\affiliation{Mitchell Institute for Fundamental Physics and Astronomy, Texas A\&M University, College Station, Texas 77843, USA}
\affiliation{Istituto Nazionale di Fisica Nucleare Trieste/Udine; $^{nn}$University of Trieste, I-34127 Trieste, Italy; $^{kk}$University of Udine, I-33100 Udine, Italy}
\affiliation{University of Tsukuba, Tsukuba, Ibaraki 305, Japan}
\affiliation{Tufts University, Medford, Massachusetts 02155, USA}
\affiliation{University of Virginia, Charlottesville, Virginia 22906, USA}
\affiliation{Waseda University, Tokyo 169, Japan}
\affiliation{Wayne State University, Detroit, Michigan 48201, USA}
\affiliation{University of Wisconsin, Madison, Wisconsin 53706, USA}
\affiliation{Yale University, New Haven, Connecticut 06520, USA}

\author{T.~Aaltonen}
\affiliation{Division of High Energy Physics, Department of Physics, University of Helsinki and Helsinki Institute of Physics, FIN-00014, Helsinki, Finland}
\author{S.~Amerio}
\affiliation{Istituto Nazionale di Fisica Nucleare, Sezione di Padova-Trento, $^{ff}$University of Padova, I-35131 Padova, Italy}
\author{D.~Amidei}
\affiliation{University of Michigan, Ann Arbor, Michigan 48109, USA}
\author{A.~Anastassov$^x$}
\affiliation{Fermi National Accelerator Laboratory, Batavia, Illinois 60510, USA}
\author{A.~Annovi}
\affiliation{Laboratori Nazionali di Frascati, Istituto Nazionale di Fisica Nucleare, I-00044 Frascati, Italy}
\author{J.~Antos}
\affiliation{Comenius University, 842 48 Bratislava, Slovakia; Institute of Experimental Physics, 040 01 Kosice, Slovakia}
\author{G.~Apollinari}
\affiliation{Fermi National Accelerator Laboratory, Batavia, Illinois 60510, USA}
\author{J.A.~Appel}
\affiliation{Fermi National Accelerator Laboratory, Batavia, Illinois 60510, USA}
\author{T.~Arisawa}
\affiliation{Waseda University, Tokyo 169, Japan}
\author{A.~Artikov}
\affiliation{Joint Institute for Nuclear Research, RU-141980 Dubna, Russia}
% ...

\collaboration{CDF Collaboration\footnote{With visitors from
$^a$University of British Columbia, Vancouver, BC V6T 1Z1, Canada,
$^b$Istituto Nazionale di Fisica Nucleare, Sezione di Cagliari, 09042 Monserrato (Cagliari), Italy,
$^c$University of California Irvine, Irvine, CA 92697, USA,
$^e$Institute of Physics, Academy of Sciences of the Czech Republic, 182~21, Czech Republic,
$^f$CERN, CH-1211 Geneva, Switzerland,
$^g$Cornell University, Ithaca, NY 14853, USA,
$^{dd}$The University of Jordan, Amman 11942, Jordan,
$^h$University of Cyprus, Nicosia CY-1678, Cyprus,
$^i$Office of Science, U.S. Department of Energy, Washington, DC 20585, USA,
$^j$University College Dublin, Dublin 4, Ireland,
$^k$ETH, 8092 Z\"{u}rich, Switzerland,
$^l$University of Fukui, Fukui City, Fukui Prefecture, Japan 910-0017,
$^m$Universidad Iberoamericana, Lomas de Santa Fe, M\'{e}xico, C.P. 01219, Distrito Federal,
$^n$University of Iowa, Iowa City, IA 52242, USA,
$^o$Kinki University, Higashi-Osaka City, Japan 577-8502,
$^p$Kansas State University, Manhattan, KS 66506, USA,
$^q$Brookhaven National Laboratory, Upton, NY 11973, USA,
$^r$University of Manchester, Manchester M13 9PL, United Kingdom,
$^s$Queen Mary, University of London, London, E1 4NS, United Kingdom,
$^t$University of Melbourne, Victoria 3010, Australia,
$^u$Muons, Inc., Batavia, IL 60510, USA,
$^v$Nagasaki Institute of Applied Science, Nagasaki 851-0193, Japan,
$^w$National Research Nuclear University, Moscow 115409, Russia,
$^x$Northwestern University, Evanston, IL 60208, USA,
$^y$University of Notre Dame, Notre Dame, IN 46556, USA,
$^z$Universidad de Oviedo, E-33007 Oviedo, Spain,
$^{aa}$CNRS-IN2P3, Paris, F-75205 France,
$^{cc}$Universidad Tecnica Federico Santa Maria, 110v Valparaiso, Chile,
$^{ll}$Universite catholique de Louvain, 1348 Louvain-La-Neuve, Belgium,
$^{oo}$University of Z\"{u}rich, 8006 Z\"{u}rich, Switzerland,
$^{pp}$Massachusetts General Hospital and Harvard Medical School, Boston, MA 02114 USA,
$^{qq}$Hampton University, Hampton, VA 23668, USA,
$^{rr}$Los Alamos National Laboratory, Los Alamos, NM 87544, USA
}}
\noaffiliation

%% file: top_dil_prd-rc_v2.bbl
\begin{thebibliography}{}

\bibitem{top_cdf}
F. Abe {\it et al.} (CDF Collaboration), {Phys. Rev. Lett.} \textbf{74}, 2626 (1995).

\bibitem{top_d0}
S. Abachi {\it et al.} (D0 Collaboration), {Phys. Rev. Lett.} \textbf{74}, 2632 (1995).

\bibitem{TOPMASS}
T. Aaltonen  {\it et al.} (CDF and D0 Collaborations), Phys.
Rev. D {\bf 86},   092003 (2012); T. Aaltonen  {\it et al.} (CDF Collaboration),
 {Phys. Rev. Lett.} \textbf{109}, 152003 (2012).

\bibitem{D0_XSEC_COMBO}
V.~M. Abazov {\it et al.} (D0 Collaboration), {Phys. Lett. B} {\bf 704}, 403 (2011). 

\bibitem{CDF_DIL_2p8}
 T. Aaltonen {\it et al.} (CDF Collaboration), {Phys. Rev. D} {\bf 82},
   052002 (2010).

\bibitem{CDF_LJ_XSEC}
  T. Aaltonen {\it et al.}  (CDF Collaboration),
  Phys.\ Rev.\ D {\bf 84}, 031101 (2011).


\bibitem{HELICITY} 
T. Aaltonen  {\it et al.} (CDF and D0 Collaborations), Phys.
Rev. D {\bf 85},  071106 (2012); T. Aaltonen  {\it et al.} (CDF Collaboration),
Phys. Rev. D {\bf 87},  031104(R) (2013).

\bibitem{RATIO}
T. Aaltonen  {\it et al.} (CDF Collaboration), {Phys. Rev. Lett.} \textbf{105},
  012001 (2010).

\bibitem{Precision_EW}
ALEPH, CDF, D0, DELPHI, L3, OPAL, SLD Collaborations, LEP Electroweak
Working Group, Tevatron Electroweak Working Group,
SLD Electroweak and Heavy Flavour Groups, arXiv:1012.2367
and references therein.

\bibitem{theory}
    M. Cacciari, S.~Frixione, M.~L. Mangano, P. Nason, and G. Ridolfi, J. High Energy Phys. 09 (2008) 127;
    N. Kidonakis and R. Vogt, Phys. Rev. D {\bf 78}, 074005 (2008);
    S. Moch and P. Uwer, Nucl. Phys. Proc. Suppl. {\bf 183}, 75 (2008);
  M. Czakon, P. Fiedler, and A. Mitov,
  arXiv:1303.6254 [hep-ph].



\bibitem{CDFII}
 R. Blair {\it et al.}   
    Fermilab-Pub-96/390-E;  D. Acosta {\em et al.} (CDF Collaboration), {Phys. Rev. D} {$\bf 71$},
032001 (2005).

\bibitem{isolated} A. Abulencia {\em et al.} (CDF Collaboration), {J. Phys. G}
\textbf{34}, 2457 (2007).

\bibitem{MET} The missing transverse energy $\met$ is defined to be
\nolinebreak{ $-\Sigma_{i}E_{T}^{i}$\^{n}$_{i}$} where $i$ is the calorimeter tower number with
$\mid\eta\mid$ $<$ 3.6, 
\^{n}$_{i}$ is a unit vector perpendicular to the beam axis and pointing at the $i$th calorimeter tower
and $E_{T}^{i}$ is the transverse energy of the $i$th tower.

\bibitem{jetclu} A. Bhatti  {\em et al.}, Nucl. Instrum. Methods Phys. Res. Sect.
A {$\bf 566$} (2006) 375.

\bibitem{secvtx} D. Acosta {\em et al.} (CDF Collaboration), {Phys. Rev. D} {$\bf 71$}, 052003 (2005).

\bibitem{pythia} T.  Sj\"ostrand, S. Mrenna, and P. Skands, {J. High Energy Phys.}
 05 (2006) 026. We use version 6.216.

\bibitem{GEANT3} GEANT3, R. Brun, F. Bruyant, M. Maire, 
A.~C. McPherson, and P. Zanarini, {C}ERN Report No.\ {C}ERN-DD/EE/84-1, (1987).

\bibitem{NLO}
  J.~M. Campbell and R.~K. Ellis, Phys. Rev. D ${\bf{60}}$,  113006 (1999).

\bibitem{PDF} A.~D.  Martin, W.~J. Stirling, R.~S. Thorne, and G. Watt {Eur.
Phys. J.  C} \textbf{63}, 189 (2009).

\bibitem{tevnphwg}The diboson cross sections when 75 $<$ $m_{\ell^{+}\ell^{-}}$ $<$ 105
GeV/$c^{2}$ are $\sigma_{\it WZ}$ = 3.22 $\pm$ 0.22~pb and $\sigma_{\it ZZ}$ = 1.20
$\pm$ 0.06~pb.  The values are calculated using \MCFM~\cite{NLO}. 
The WZ and ZZ cross sections given in the main text are larger,
because in the corresponding samples the dilepton mass phase space is restricted only to 
$m_{\ell^{+}\ell^{-}} > $ 2 GeV/$c^2$. This allows for the appearance of off-shell Z bosons in
the samples, which contribute negligibly to the top dilepton background but can
be an important component of other CDF analyses, with which these samples are shared.
 

\bibitem{alpgen} M.~L. Mangano, M. Moretti, F. Piccinini, R. Pittau, and A.~D.
Polosa, {J. High Energy Phys.} 07 (2003) 001. We use version 2.10.

\bibitem{baur}
U. Baur and E.~L. Berger, Phys. Rev. D ${\bf{41}}$, 1476 (1990).

\bibitem{baur2}
U. Baur, T. Han, and J. Ohnemus, Phys. Rev. D ${\bf{48}}$, 5140 (1993). 

\bibitem{JES}
A. Abulencia {\it et al.} (CDF Collaboration),  {Phys. Rev. D} {$\bf 73$},
  032003 (2006); T. Aaltonen  {\it et al.} (CDF Collaboration), {Phys. Rev. D}
 {$\bf 81$},  052011 (2010).

\bibitem{lum} D. Acosta {\em  et al.}, {Nucl. Instrum. Methods Phys. Res. Sect. A}
\textbf{494}, 57 (2002).



\end{thebibliography}
